# Ground State Calculations of the Confined Molecular Ions $H_2^+$ and $HeH^{++}$ Using Variational Monte Carlo Method

S. B. Doma[1], F. N. El-Gammal[2] and A. A. Amer[2]

[1] Department of Mathematics and Computer Science, Faculty of Science, Alexandria University, Alexandria, Egypt
E-mail address: sbdoma@yahoo.com
[2] Mathematics Department, Faculty of Science, Menofia University, Shebin El-Kom, Egypt

**Absract**

The ground state energy of hydrogen molecular ion $H_2^+$ confined by a hard prolate spheroidal cavity is calculated. The case in which the nuclear positions are clamped at the foci is considered. Our calculations are based on using the variational Monte Carlo method with an accurate trial wave function depending on many variational parameters. The calculations were extended also to include the $HeH^{++}$ molecular ion. The obtained results are in good agreement with the recent results.

## 1- Introduction

Recently, studies of quantum objects confined by different forms of external potentials have attracted the attention of both physicists and quantum chemists. This is due to the unusual physical and chemical properties observed in such systems when submitted to narrow spatial limitation as compared to their free cases. Also, confined systems are widely used to model a variety of problems in physics and chemistry. Examples of these problems occur in the study of the synthesis of nanostructure materials such as carbon nanotubes [1,2], buckyballs and zeolitic nanochannels which serve as ideal containers for molecular insertion and storage with promising applications [3–6]. The increasing pace at which research is being carried out in the aforementioned systems demands many powerful and sophisticated methodologies (Hartree–Fock, quantum chemical density functional theory, quantum molecular dynamics, to mention a few [1, 7–9]) and also, complementary exploratory models aimed at understanding the basic mechanisms of the changes in the electronic and structural properties of confined molecules. Various theoretical models have been proposed in the past to analyze the confinement effects on the confined systems, particularly those based on boxed-in molecules. Box models of confinement with hard and soft boundaries have been widely used to survey the effect of spatial limitation in the case of simple molecules such as $H_2^+$ molecular ion, $H_2$ molecule and some small polyatomics as $H_2O$, $CH_4$, $NH_3$ and $LiF$. Since, the $H_2^+$ molecular ion is considered as one of the first non-trivial quantum mechanical systems, many studies have been presented to investigate the effect of compression on its properties. In Ref [10], Molinar-Tabares *et al.* studied the $H_2^+$ confined by spheroids of size $\xi_0$ using prolate spheroidal coordinates. In frame of Born–Oppenheimer approximation, the Schrödinger equation was solved by the method of separation of variables to obtain the equilibrium distance between nuclei and the corresponding energy as a function of $\xi_0$. Also, the vibrational energy of the nuclei, the pressure, the polarizability and the anisotropy were calculated.

A first successful attempt to uncouple the nuclear positions from the foci was made by Crus *et al.* [11] for $H_2^+$ confined within impenetrable prolate spheroidal boxes. The non-separable Schrödinger problem was solved using the variational method with simple LCAO Dickinson type variational ansatz wave function to obtain the ground state energies of the enclosed $H_2^+$ and $HeH^{++}$ when the nuclear positions do not coincide with the foci. The obtained results were the first of this kind. The pervious results were extended to cover the case of $H_2$ by Colin-Rodriguez *et al.* [12]. It was shown that by making the cavity size and shape independent of the nuclear positions, optimum equilibrium bond lengths and energies are obtained as compared with corresponding on-focus calculations. This procedure allows for a controlled treatment of molecular properties by selecting an arbitrary size and shape of the confining spheroidal box. A generalization of previous theoretical studies of molecular confinement based on the molecule-in-a-box model was presented in Ref [13]. The non-separable Schrödinger problem for the confined $H_2^+$ and $H_2$ molecules in their ground states is treated variationally. In both cases, it was shown that, the equilibrium bond length and total energy depend significantly on the confining-barrier height for fixed cavity sizes and shapes. Also, this study shows that as the cavity size is reduced, the limit of stability of the confined molecule is attained for a critical size. One of the most and recent studies was presented by Sarsa *et al.* [14] where the $H_2^+$ molecular ion confined by impenetrable spherical surfaces was studied beyond the Born–Oppenheimer approximation. The confinement of both electron and nuclei were considered and they could show that the electron constraint is much more efficient to increase the energy than the nuclei confinement. This study shows that a metastable bound state of the ion, above the energy corresponding to the dissociation limit, can be obtained when the electron constraint is present.

In the present paper we extended our previous works about the applications of the variational Monte Carlo (VMC) method to the atomic systems [15-19], to the investigation of the molecular systems. As a first application we studied the confined hydrogen molecular ion $H_2^+$. Accordingly, we considered the confined $H_2^+$ placed inside spherical hard boxes in framework of the VMC method, which was used before to study the $H_2^+$ in the unconfined case by Alexander *et al.* [20]. Our calculations were extended also to include the $HeH^{++}$ molecular ion.

## 2- Method of the Calculations

Our calculations are based on using the VMC method which is considered as a one of the most important Quantum Monte Carlo methods. It is based on a combination of two ideas namely the variational principle and the Monte Carlo evaluation of integrals using importance sampling based on the Metropolis algorithm [21]. The VMC methods are used to compute quantum expectation values of an operator with a given trial wave function. Given a Hamiltonian $H$ and a trial wave function $\psi_T$, the variational principle states that the expectation value is the varitional energy [22]:

$$E_{VMC} = \frac{\int \psi_T^*(R) H \psi_T(R) dR}{\int \psi_T^*(R) \psi_T(R) dR} \geq E_{exact}, \qquad (2.1)$$

where $\psi_T$ is a trial wave function, $R$ is the $3N$-dimensional vector of the electron coordinates and $E_{exact}$ is the exact value of the energy of that state. Also, it is important to calculate the standard deviation of the energy [22]

$$\sigma = \sqrt{\frac{\langle E_L^2 \rangle - \langle E_L \rangle^2}{(N-1)}} \qquad (2.2)$$

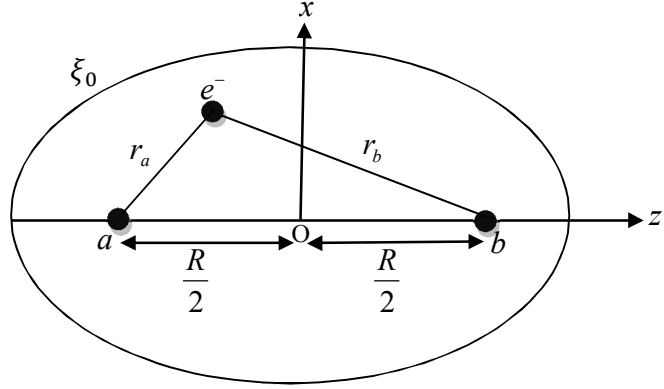

Fig.1 Confined hydrogen molecular ion $H_2^+$, confined within a prolate spheroidal cavity.

**3- The Hamiltonian of the System and the Trial Wave Function**
We consider the $H_2^+$ molecular ion confined within a prolate spheroidal cavity defined by the geometric contour $\xi_0$ and characterized by a barrier potential of height $V$, as shown in Figure-1. The Schrödinger equation for the confined $H_2^+$ can be written as follows [14]

$$H\psi(\mathbf{r}_e, \mathbf{R}_1, \mathbf{R}_2) = E\psi(\mathbf{r}_e, \mathbf{R}_1, \mathbf{R}_2), \qquad (3.1)$$

where, $\mathbf{r}_e$ is the position vector of the electron, $\mathbf{R}_1$ and $\mathbf{R}_2$ are the position vectors of the nuclei.

Considering the origin is placed at the center of mass of the nuclei, then the molecular Hamiltonian for the confined $H_2^+$ molecular ion may be conveniently written (in atomic units) as follows [14]

$$H = -\frac{1}{2\varepsilon}\nabla_{r_e}^2 - \frac{1}{2\upsilon}\nabla_R^2 - \frac{Z_a}{r_a} - \frac{Z_b}{r_b} + \frac{Z_a Z_b}{R} + V_c, \qquad (3.2)$$

where, $R$ is the internuclear separation, $R = |\mathbf{R}_2 - \mathbf{R}_1|$. $V_c$ is the confining potential barrier imposed by the spheroidal boundary, $S$, which in our case is assumed to be infinitely high, that is

$$V_c = \begin{cases} \infty & (r_a, r_b) \in S \\ 0 & (r_a, r_b) \notin S \end{cases} \qquad (3.3)$$

In the above equation, $r_a$ and $r_b$ denote the distances from the electron to nuclei $a$ and $b$

$$r_a = \sqrt{x^2 + y^2 + \left(z + \tfrac{1}{2}R\right)^2}, \qquad r_b = \sqrt{x^2 + y^2 + \left(z - \tfrac{1}{2}R\right)^2}, \qquad (3.4)$$

The notations $\varepsilon$ and $\upsilon$ stand for the reduced masses and defined as

$$\varepsilon = \frac{m(M_1 + M_2)}{m + M_1 + M_2}, \qquad \upsilon = \frac{M_1 M_2}{M_1 + M_2}. \qquad (3.5)$$

In this paper we will consider the case of prolate spheroidal confining box, so we will use prolate spheroidal coordinates. In geometry, the elliptic coordinate system consists of families of mutually orthogonal confocal ellipsoids ($\lambda$) and hyperboloids ($\mu$) of revolution. In addition, Schrödinger's equation can be solved for the free and confined $H_2^+$ ion on condition that the nuclear positions coincide with the foci [23-25]. The prolate spheroidal coordinates are defined by [12]

$$\lambda = \frac{r_a + r_b}{R}, \qquad \mu = \frac{r_a - r_b}{R}, \qquad \varphi = \varphi \ (\varphi \text{ is the azimuthal angle}). \tag{3.6}$$

The ranges of these variables are

$$1 \leq \lambda \leq \infty, \qquad -1 \leq \mu \leq 1, \qquad 0 \leq \varphi \leq 2\pi. \tag{3.7}$$

In this coordinates, the kinetic energy operator is written as [26]

$$-\frac{1}{2}\nabla_{r_e}^2 = -\frac{2}{R^2(\lambda^2-\mu^2)}\left\{\frac{\partial}{\partial\lambda}(\lambda^2-1)\frac{\partial}{\partial\lambda} + \frac{\partial}{\partial\mu}(1-\mu^2)\frac{\partial}{\partial\mu} + \frac{(\lambda^2-\mu^2)}{(\lambda^2-1)(1-\mu^2)}\frac{\partial^2}{\partial\varphi^2}\right\}. \tag{3.8}$$

In Fig.1 we present the geometric characteristics of the confined hydrogen molecular ion $H_2^+$ confined within a prolate spheroidal cavity defined by $\lambda = \xi_0$. The nuclear charges $Z_a = Z_b$ are both located at distance $\frac{R}{2}$ from the origin. From the figure it was easy to write down the Hamiltonian operator corresponding to the coordinates of the electron and the two nuclei. Because the spheroidal coordinates depend on $R$, the effect on the nuclear radial derivatives of transforming to spheroidal coordinates must be considered. At this point, these derivatives are evaluated holding the body-frame Cartesian coordinates fixed. It is preferable, however, to evaluate them holding the spheroidal coordinates fixed instead.

Using the chain rule, the radial part of the nuclear Laplacian can be written as [27]

$$\nabla_R^2 = \left(\frac{\partial^2}{\partial R^2} + \frac{2}{R}\frac{\partial}{\partial R}\right)_{xyz} = \left(\frac{\partial^2}{\partial R^2}\right)_{\lambda\mu\varphi} - \left(\frac{2\left(Y+\frac{3}{2}\right)}{R}\frac{\partial}{\partial R}\right)_{\lambda\mu\varphi} + \frac{\left(Y+\frac{3}{2}\right)\left(Y+\frac{5}{2}\right)}{R^2}, \tag{3.9}$$

keeping in mind that the radial derivatives are taken holding $\lambda\mu\varphi$ fixed. The operator $Y$ in Eq. (3.9) involves only electronic coordinates and is defined as [27]

$$Y = \frac{1}{(\lambda^2-\mu^2)}\left\{(\lambda+\gamma\mu)(\lambda^2-1)\frac{\partial}{\partial\lambda} + (\mu+\gamma)(1-\mu^2)\frac{\partial}{\partial\mu}\right\}, \tag{3.10}$$

where $\gamma$ is the mass asymmetry parameter defined as [27]

$$\gamma = \frac{M_1 - M_2}{M_1 + M_2}. \tag{3.11}$$

For homonuclear systems, this definition corresponds precisely to the standard definition of $Y$, since the mass asymmetry parameter $\gamma$ is then zero.
[For the $H_2^+$, we used $M_1 = M_2 = 938.2720$ MeV c$^{-2}$, then the parameter $\gamma$ is equal to zero.]

The used trial wave function for solving the Schrödinger equation of confined $H_2^+$ in our calculations is proposed by Ishikawa *et al.* [26]. This trial wave function depends on the Slater type function $\psi_0$ as initial function which takes the form:

$$\psi_0 = exp(-\omega\lambda), \tag{3.12}$$

where $\omega$ is nonlinear parameter. In this choice, the trial wave function $\psi$ is generated in the analytical expansion form of

$$\psi = \sum_i C_i \lambda^{m_i} \mu^{n_i} exp(-\omega\lambda), \qquad (3.13)$$

where $C_i$ are the variational parameters and $m_i$ are positive or negative integers. Since the $1s\sigma_g$ ground-state has a gerade symmetry, $n_i$ should be zero or a positive even integer. Applying the iterative complement interaction (ICI) method, Ishikawa *et al.* [26] used this trial wave function to calculate energies for the ground-state $1s\sigma_g$ of $H_2^+$ at different orders and the first excited state $1s\sigma_u$ (ungerade) in the free (unconfined) case. They could obtain very accurate results compared to the corresponding exact values.

For the ground state, the overlap and Hamiltonian integrals of $H_2^+$ are easily done when the wave function is given by Eq. (3.13). To study the case of $H_2^+$ confined by a hard spherical boundary surface, the wave function must vanishes at the spherical boundary surface, so a cut-off factors is employed to fulfill this condition and the wave function becomes:

$$\psi = \begin{cases} \sum_i C_i \lambda^{m_i} \mu^{n_i} exp(-\alpha\lambda) \times \left[\left(1 - \frac{\lambda-1}{\xi_0-1}\right) exp\left(\frac{\lambda-1}{\xi_0-1}\right)\right] & \text{for } \lambda < \xi_0 \\ 0 & \text{for } \lambda \geq \xi_0 \end{cases}. \qquad (3.14)$$

In Eq. (3.14) the last factor in parenthesis represents the cut-off factor in terms of the elliptic coordinates and it guarantee that $\psi(\lambda = \xi_0, \mu) = 0$ at the boundary. This type of cut-off function was introduced in [28] and it was found to provide accurate results.

**4- Results and Discussion**

The variational Monte Carlo method has been employed for the ground state of free and confined $H_2^+$. For the $H_2^+$, we used $M_1 = M_2 = 938.2720$ MeV c$^{-2}$ and $m = 0.5109989$ MeV c$^{-2}$ obtaining $\varepsilon = 0.9997278$ and $\upsilon = 918.0763$ in atomic units of mass. The hydrogen molecular ion $H_2^+$ has the charge parameters $Z_a = Z_b = 1$. All energies are obtained in atomic units i.e. ($\hbar = e = m_e = 1$) with set of $4 \times 10^7$ Monte Carlo integration points in order to make the statistical error as low as possible. To gain some confidence on the adequacy of the trial wave function given by Eq. (3.14) for our calculations in frame of VMC method, we first consider the case of unconfined ion for different three values. In Table-1 we compare the predictions of our results for the behavior of the total energy of the ground-state ($1s\sigma_g$) of free $H_2^+$ ion at different values for the internuclear distance $R$ with the exact calculations by Wind [29]. Excellent quantitative agreement is obtained compared to the corresponding exact values. These results validate the accuracy for the wave function to calculate the ground-state energy ($1s\sigma_g$) of $H_2^+$ under compression. Since molecules when squeezed into a tiny space, present different electronic and structural behavior in contrast to their free condition; then, knowledge of the way these changes take place as a function of cavity size, shape and composition is of paramount importance. Firstly, we will study the case in which the nuclei are clamped at the foci and interfocal distance $R = 2$ a.u., together with the other approximate calculations. The potential barrier parameter $V_c$ can take values between zero and infinity representing walls with increasing confining strength. In Table-2 we displayed the results obtained for the ground-state of the confined $H_2^+$ molecular ion together with the corresponding results available in the literature and the most recent results. The obtained energies were calculated for wide range of $\xi_0$. The small values of $\xi_0$ describe the case of strong confinement where the large values represent the weak compression. It is clear that our results are of good agreement in comparison with previous data. The agreement with other data is found to be good even for relatively large values of the eccentricity $1/\xi_0$.

Table-1 The electronic energy of the unconfined $H_2^+$ ion for various internuclear distance $R$ compared with the exact values. In parentheses, we show the statistical error in the last figure.

| $R$ | $E_{this\ work}$ | $E_{exact}^a$ |
|---|---|---|
| 1.0 | -1.455 911(6) | -1.451 786 313 378 1 |
| 2.0 | -1.102 634(5) | -1.102 634 214 494 9 |
| 3.0 | -0.910 832(3) | -0.910 896 197 382 3 |

$^a$Ref [29].

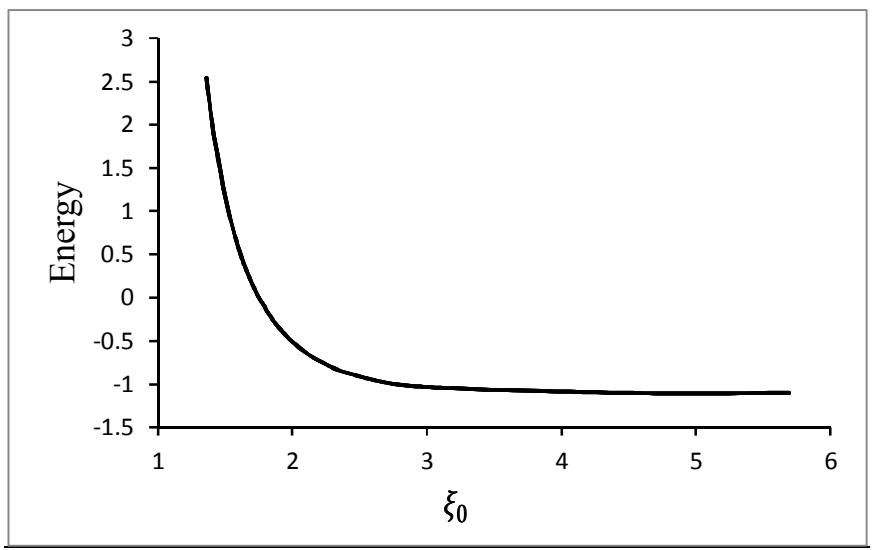

Figure-2 The ground-state energy of $H_2^+$ versus $\xi_0$.

Figure-2 shows the variation of the ground-state energy with respect to $\xi_0$. It is clear from Figure-2 that the energies increase when $\xi_0$ decreases for strong compression at $\xi_0 < 2.9162$ where for large values of $\xi_0 \geq 2.9162$, the compression effect becomes not noticeable and the energy is nearly stable and approaches to the corresponding exact value, i.e. when $\xi_0$ increases that leads to less energy to reach even the free state of $H_2^+$. Also, Figure-2 insures the fact that the energy of the low-lying states in a confined quantum charged system is determined by a competition of confinement kinetic energy and Coulomb interaction energy. As the molecules are compressed, they become constrained in a diminishing spherical box so that according to the quantum mechanical uncertainty principle, the electrons increase their momentum and thereby leading to a net gathering of kinetic energy. In other meaning the smaller the confined potential spheroidal cavity $\xi_0$ is, the higher the confinement kinetic energy is. When the increase in the confinement kinetic energy becomes predominant and cannot be compensated by the increase of the Coulomb attractive energy, the energies of the confined $H_2^+$ increase.

On the other hand, Table-3 displays the results for the energy evolution of the ($1s\sigma_g$) state of the confined $H_2^+$ molecular ion within a prolate spheroidal cavity with fixed major axis $C = R\xi_0 = 5$ a.u. and different values for the internuclear distance $R$, compared to the corresponding exact calculations by Mateos *et al.* [25] and other approximate calculations. The agreement with other data is found to be good.

Table-2 Energy of the ground state of $H_2^+$ obtained using the wave function of Eq. (3.14) with a fixed internuclear distance $R = R_0 = 2$ a.u and different sizes and eccentricities $\xi_0$ as compared with the exact and other approximate calculations. In parentheses, we show the statistical error in the last figure.

| $\xi_0$ | $E_{\text{this work}}$ | $E^a$ | $E^b$ | $E^c$ | $E^d_{\text{exact}}$ |
|---|---|---|---|---|---|
| 5.6924 | -1.102509(7) | -1.1022 | -1.1021 | - | -1.1025 |
| 4.4468 | -1.099999(1) | - | - | -1.099991 | -1.1 |
| 2.9162 | -1.025060(1) | - | -1.0236 | - | -1.025 |
| 2.4196 | -0.875086(3) | - | -0.8745 | -0.875027 | -0.875 |
| 2.2237 | -0.750413(5) | -0.7499 | -0.749 | - | -0.75 |
| 2.0917 | -0.625237(6) | - | - | -0.624975 | -0.625 |
| 1.9934 | -0.499707(8) | -0.4999 | -0.49 | - | -0.5 |
| 1.9002 | -0.347878(2) | - | - | -0.3467505 | -0.35 |
| 1.8638 | -0.274803(2) | - | - | - | -0.275 |
| 1.8186 | -0.174838(2) | - | - | -0.1750125 | -0.175 |
| 1.7788 | -0.072020(3) | - | - | - | -0.075 |
| 1.7606 | -0.024874(4) | - | - | - | -0.025 |
| 1.7434 | 0.0247(2) | 0.0258 | - | - | 0.025 |
| 1.7270 | 0.074663(4) | - | - | - | 0.075 |
| 1.7115 | 0.121605(3) | - | - | - | 0.125 |
| 1.6690 | 0.2745(2) | - | - | - | 0.275 |
| 1.6150 | 0.5012(3) | 0.5025 | 0.507 | - | 0.5 |
| 1.5229 | 1.0095(4) | - | - | - | 1.0 |
| 1.4555 | 1.541445(1) | - | - | - | 1.5 |
| 1.4035 | 2.015272(1) | - | - | - | 2.0 |
| 1.3621 | 2.544751(1) | 2.5214 | - | - | 2.5 |

$^a$Ref [11].    $^b$Ref [14].    $^c$Ref [30].    $^d$Ref [24].

Finally, our results were extended to include the $HeH^{++}$ molecular ion which has the charge parameters $Z_a = 2$, $Z_b = 1$. Table-4 shows the results for the energy evolution of the ($1s\sigma_g$) state of the confined $HeH^{++}$ molecular ion confined by a hard prolate spheroid characterized by an internuclear distance $R = 2$ a.u. with different sizes and eccentricities. Also, the corresponding exact calculations by Ley-Koo *et al.* [24] and accurate variational calculations from Ref [11] are presented for comparison. The comparison insures that our results are of good accuracy. It is clear that the obtained numerical results are in good agreement with the exact and other approximate calculations.

Table-3 Total energy behavior of the ground state energy of $H_2^+$ enclosed by a prolate spheroidal cavity with major axis $C = R\xi_0 = 5$ a.u. and varying internuclear distances. In parentheses, we show the statistical error in the last figure.

| $R$ | $E_{\text{this work}}$ | $E^a$ | $E^b_{\text{exact}}$ |
|---|---|---|---|
| 1.1 | -0.4190592(5) | -0.4287 | -0.429173 |
| 1.4 | -0.4739613(3) | -0.4716 | -0.471751 |
| 1.5 | -0.4710616(2) | -0.4703 | -0.471784 |
| 1.6 | -0.4657539(1) | -0.4657 | -0.466979 |
| 1.9 | -0.4302683(4) | -0.4294 | -0.430244 |
| 2.2 | -0.3669057(2) | -0.3665 | -0.366949 |
| 2.5 | -0.2791488(7) | -0.2789 | -0.279130 |
| 2.8 | -0.1631775(1) | -0.1634 | -0.163430 |

$^a$Ref [11].    $^b$Ref [25].

Table-4 The electronic energy of the $(1s\sigma_g)$ of $HeH^{++}$ with nuclear positions located at the foci of a confining prolate spheroidal cavity of internuclear distance $R = 2$ a.u. with different sizes and eccentricities $\xi_0$. In parentheses, we show the statistical error in the last figure.

| $\xi_0$ | $E_{\text{this work}}$ | $E^a$ | $E^b_{\text{exact}}$ |
|---|---|---|---|
| 1.7025 | -1.501766(8) | - | -1.5 |
| 1.6580 | -1.324248(6) | - | -1.325 |
| 1.6410 | -1.250694(8) | -1.2498 | -1.25 |
| 1.5914 | -1.000004(1) | - | -1.0 |
| 1.5499 | -0.7501661(3) | -0.7498 | -0.75 |
| 1.5424 | -0.700063(3) | - | -0.7 |
| 1.5351 | -0.6505256(4) | - | -0.65 |
| 1.5211 | -0.5502026(5) | -0.5498 | -0.55 |
| 1.5144 | -0.5009148(4) | - | -0.5 |
| 1.4833 | -0.2506181(1) | - | -0.25 |
| 1.4558 | 0.000497736(7) | - | 0.0 |
| 1.4313 | 0.2515458(8) | 0.2519 | 0.25 |
| 1.4091 | 0.5058985(1) | - | 0.5 |
| 1.3705 | 1.056358(5) | 1.0066 | 1.0 |
| 1.3379 | 1.502523(5) | - | 1.5 |

$^a$Ref [11].    $^b$Ref [24].

**5- Conclusion**

In this paper we have used the variational Monte Carlo method to study $H_2^+$ molecular ion confined by impenetrable spherical surfaces. We have calculated the energies for both confined $H_2^+$ molecular ion and its free case. We considered the case of small values of $\xi_0$ which describe the strong compression as well as the case of large values of $\xi_0$. Our results were extended also to include the $HeH^{++}$ molecular ion. The energy was plotted as a function of $\xi_0$ to show graphically the effect of compression on the total energy. The graphs indicate that the values of the energy affected significantly at small values of $\xi_0$, where at large values the energy tends to be constant and approaches to its uncompressed value. In both cases our results exhibit good accuracy comparing with previous values obtained by using different methods and different forms of trial wave functions. Finally, we conclude that the applications of VMC method can be extended successfully to cover the case of compressed molecules.